\documentclass[prl,superscriptaddress,preprint]{revtex4}

\newcommand{\beq}{\begin{equation}}
\newcommand{\eeq}{\end{equation}}
\newcommand{\barr}{\begin{eqnarray}}
\newcommand{\earr}{\end{eqnarray}}

\usepackage[ansinew]{inputenc}
\usepackage{graphicx}
\usepackage{amsmath}
\usepackage{amsthm}
\usepackage{bm}
\usepackage{layout}
\usepackage{float}
\usepackage{txfonts}
\usepackage{amsfonts}
\usepackage{amssymb}%

\begin{document}


\title{Multidimensional hydrogenic complexity}

\author{Sheila L\'opez-Rosa}
\email{slopez@ugr.es}
\affiliation{Departamento de F\'isica At\'omica, Molecular y Nuclear, Universidad de Granada, 18071 Granada,
Spain}
\affiliation{Instituto Carlos I de F\'isica Te\'orica y Computacional, Universidad de Granada,
18071 Granada, Spain}

\author{Daniel Manzano}
\email{manzano@ugr.es}
\affiliation{Departamento de F\'isica At\'omica, Molecular y Nuclear, Universidad de Granada, 18071 Granada,
Spain}
\affiliation{Instituto Carlos I de F\'isica Te\'orica y Computacional, Universidad de Granada,
18071 Granada, Spain}

\author{Jes\'us S. Dehesa}
\email{dehesa@ugr.es}
\affiliation{Departamento de F\'isica At\'omica, Molecular y Nuclear, Universidad de Granada, 18071 Granada,
Spain}
\affiliation{Instituto Carlos I de F\'isica Te\'orica y Computacional, Universidad de Granada,
18071 Granada, Spain}


\begin{abstract}
The internal disorder of a D-dimensional hydrogenic system, which is strongly associated to the non-uniformity of the quantum-mechanical density of its physical states, is investigated by means of the shape complexity in the two reciprocal spaces. This quantity, which is the product of the disequilibrium and the Shannon entropic power, is mathematically expressed for both ground and excited stationary states in terms of certain entropic functionals of Laguerre and Gegenbauer (or ultraspherical) polynomials. Emphasis is made in the ground and circular states.

\end{abstract}

\maketitle

\section{Introduction}

The hydrogenic system (i.e., a negatively-charged particle moving around a positively-charged core which electromagnetically binds it in its orbit) with dimensionality $D\geq 1$, plays a central role in D-dimensional quantum physics and chemistry \cite{H93, A00}. It includes not only a large variety of three-dimensional physical systems (e.g., hydrogenic atoms and ions, exotic atoms, antimatter atoms, Rydberg atoms,\dots) but also a number of nanoobjects so much useful in semiconductor nanostructures (e.g., quantum wells, wires and dots) \cite{H05, L07} and quantum computation (e.g., qubits) \cite{N00, D03}. Moreover it has a particular relevance for the dimensional scaling approach in atomic and molecular physics \cite{H93} as well as in quantum field theory \cite{I06}. Let us also say that the existence of hydrogenic systems with non standard dimensionalities has been shown for $D<3$ \cite{L07} and suggested for $D>3$ \cite{B99}.

The internal disorder of this system, which is manifest in the non-uniformity quantum-mechanical density and in the so distinctive hierarchy of its physical states, is being increasingly investigated beyond the root-mean-square or standard deviation (also called Heisenberg measure) by various information-theoretic elements; first, by means of the Shannon entropy \cite{Y94, D98, D01} and then, by other individual information and/or spreading measures as the Fisher information and the power and logarithmic moments ( see \cite{V00}, and   especially \cite{D08}). Just recently, further complementary insights have been shown to be obtained in the three-dimensional hydrogen atom by means of composite information-theoretic measures, such as the Fisher-Shannon and the shape complexity \cite{S08, L08}. In particular, Sa\~nudo and Lopez-Ruiz \cite{S08} have found some numerical evidence that, contrary to the energy, both the Fisher-Shannon measure and the shape complexity in the positi
 on space do not present any accidental degeneracy (i.e. they do depend on the orbital quantum number $l$); moreover, they take on their minimal values at the circular states (i.e., those with the highest $l$). In fact, the position Shannon entropy by itself has also these two characteristics as it has been numerically pointed out long ago \cite{Y94}, where the dependence on the magnetic quantum number is additionally studied for various physical states.

The shape complexity \cite{C02} occupies a very especial position not only among  the composite information-theoretic measures in general, but also within the class of measures of complexity. This is because of the following properties: (i) invariance under replication, translation and rescaling transformations, (ii) minimal value for the simplest probability densities (namely, e.g. uniform and Dirac's delta in one-dimensional case), and (iii) simple mathematical structure: it is given as the product of the disequilibrium or averaging density and the Shannon entropy power of the system.

In this work we provide the analytical methodology to calculate the shape complexity of the stationary states of the  D-dimensional hydrogenic system in the two reciprocal position and momentum spaces and later we apply it to a special class of physical states which includes the ground state and the circular states (i.e. states with the highest hyperangular momenta allowed within a given electronic manifold). First, in Section 2, we briefly describe the known expressions of the quantum-mechanical density of the system in both spaces. In Section 3 we show that the computation of the two shape complexities for arbitrary D-dimensional hydrogenic stationary states boils down to the evaluation of some entropic functionals of Laguerre and Gegenbauer polynomials. To have the final expressions of these complexity measures in terms of the dimensionality D and the quantum numbers characterizing the physical state under consideration, we need to compute the values of these polynomial en
 tropic functionals what is, in general, a formidable open task. However, in Section 4, we succeed to do it for the important cases of ground and circular states. It seems that for the latter ones the shape complexity has the minimal values, at least in the three-dimensional case as indicated above. It is also shown that our results always fulfil the uncertainty relation satisfied by the position and momentum shape complexities \cite{L08}. Finally, some conclusions are given.

\section{The D-dimensional hydrogenic quantum-mechanical densities}\label{sec2}

Let us consider an electron moving in the D-dimensional Coulomb ($D \ge 2$) potential $V(\vec{r})=-\frac{Z}{r}$, where $\vec{r}=(r, \theta_1,\theta_2,...,\theta_{D-1})$ denotes the electronic vector position in polar coodinates. The stationary states of this hydrogenic system are described by the wavefunctions
\beq
\Psi_{n,l,\left\lbrace \mu \right\rbrace }\left(\vec{r},t \right)=\psi_{n,l, \left\lbrace \mu \right\rbrace }\left(\vec{r}\right) \exp \left(- i E_n t\right),
\nonumber
\eeq
where $\left(E_n,\Psi_{n,l, \left\lbrace \mu \right\rbrace }  \right)$ denote the physical solutions of the Schr\"odinger equation of the system \cite{H93, A00, D08}. Atomic units $\left(\hslash=e=m_e=1 \right)$ are used throughout the paper. The energies are given by
\beq 
E= -\frac{Z^2}{2\eta^2},\hspace{0.5cm} \text{ with } \hspace{0.5cm} \eta=n+\frac{D-3}{2}; \hspace{5mm} n=1,2,3,...,
\label{energia}
\eeq
and the eigenfunctions can be expressed as
\beq
\Psi_{n,l, \left\lbrace \mu \right\rbrace }(\vec{r})=R_{n,l}(r) {\cal{Y}}_{l,\{\mu\}}(\Omega_{D-1}),
\label{FunOndaR}
\eeq
where $(l,\left\lbrace \mu \right\rbrace)\equiv(l\equiv\mu_1,\mu_2,...,\mu_{D-1})\equiv(l,\{\mu\})$ denote the hyperquantum numbers associated to the angular variables $\Omega_{D-1}\equiv (\theta_1, \theta_2,...,\theta_{D-1} \equiv \varphi)$, which may have all values consistent with the inequalities $l\equiv\mu_1\geq\mu_2\geq...\geq \left|\mu_{D-1}\right|\equiv \left|m\right|\geq 0$. The radial function is given by
\beq
R_{n,l}(r)= \left( \frac{\lambda^{-D}}{2 \eta}\right)^{1/2}   \left[\frac{\omega_{2L+1}(\hat{r})}{\hat{r}^{D-2}}\right]^{1/2}{\tilde{\cal{L}}}_{\eta-L-1}^{2L+1}(\hat{r}),
\label{radial}
\eeq
where ${\tilde{\cal{L}}}_{k}^{\alpha}(x)$ denotes the Laguerre polynomials of degree $k$ and parameter $\alpha$, orthonormal with respect to the weight function $\omega_\alpha (x)= x^\alpha e^{-x}$, and the grand orbital angular momentum hyperquantum number $L$ and the adimensional parameter $\hat{r}$ are
\beq 
L=l+\frac{D-3}{2}, \hspace{0.5cm} l=0,1,2,... \hspace{0.5cm} \text{ and } \hspace{0.5cm} \hat{r}=\frac{r}{\lambda},\hspace{0.5cm} \text{ with } \hspace{0.5cm}\lambda=\frac{\eta}{2Z}.
\label{nL}
\eeq

The angular part ${\cal{Y}}_{l,\{\mu\}}(\Omega_{D-1})$ is given by the hyperspherical harmonics \cite{A00,A93}
\beq
{\cal{Y}}_{l,\{\mu\}}(\Omega_{D-1})=\frac{1}{\sqrt{2 \pi}} e^{im\varphi} \prod^{D-2}_{j=1} {\tilde{C}}_{\mu_{j}-\mu_{j+1}}^{\alpha_j+\mu_{j+1}}(\cos \theta_j) \left( \sin \theta_j\right)^{\mu_{j+1}},
\label{hyperesf}
\eeq
with $\alpha_j= \frac{1}{2} (D-j-1)$ and ${\tilde{C}}^{\lambda}_{k}(x)$ denotes the orthonormal Gegenbauer polynomials of degree $k$ and parameter $\lambda$.

Then, the quantum-mechanical probability density of the system in position space is
\beq
\rho_{n,l,\left\lbrace \mu \right\rbrace} (\vec{r}) = \left|\Psi_{n,l,\left\lbrace \mu \right\rbrace} \left( \vec{r} \right)  \right|^2 = R_{n,l}^{2} (r) \left|{\cal{Y}}_{l,\left\lbrace \mu \right\rbrace} \left( \Omega_{D-1}\right)  \right|^2 ,
\label{probdif}
\eeq

In momentum space the eigenfunction of the system is \cite{A00, D08, DL07, AQ}
\beq
\tilde{\Psi}_{nl\{\mu\}}(\vec{p})={\cal{M}}_{n,l}(p){\cal{Y}}_{l\{\mu\}}(\Omega_{D-1}),
\label{FunOndaP}
\eeq
where the radial part is
\begin{align}\label{momento}
{\cal{M}}_{n.l}(p)&= 2^{L+2} \left(  \frac{\eta} {Z}\right) ^{D/2}  \frac{(\eta \tilde{p})^l}{(1+\eta^2 {\tilde{p}}^2)^{L+2}} {\tilde{C}}^{L+1}_{\eta-L-1}\left(\frac{1-\eta^2 \tilde{p}^2}{1+\eta^2 \tilde{p}^2}\right)\\
&=\left(  \frac{\eta} {Z}\right) ^{D/2} (1+y)^{3/2} \left(\frac{1+y}{1-y} \right)^{\frac{D-2}{4}} {\omega^{*}}^{1/2}_{L+1} (y) {\tilde{C}}^{L+1}_{\eta-L-1}(y),\nonumber
\end{align}
with $y=\frac{1-\eta^2 \tilde{p}^2 }{1+\eta^2 \tilde{p}^2}$ and $\tilde{p}=\frac{p}{Z}$ (here the electron momentum $p$ is assumed to be expressed in units of $p_\mu$, where $p_{\mu_r}=\frac{\mu_r}{m_e}p_0=\mu_r$ m.a.u, since $m_e=1$ and the momentum atomic unit is $p_0=\frac{\hbar}{a_0}=\frac{m_e e^2}{\hbar}$; $\mu_r$ is the reduced mass of the system). The symbol ${\tilde{C}}^\alpha_m(x)$ denotes the Gegenbauer polynomial of order $k$ and parameter $\alpha$ orthonormal with respect to the weight funciont $\omega^{*}_{\alpha}=(1-x^2)^{\alpha-\frac{1}{2}}$ on the interval $\left[-1,+1 \right]$. The angular part is again an hyperspherical harmonic as in the position case, but with the angular variables of the vector $\vec{p}$. Then, one has the following expression
\beq
\gamma(\vec{p})=\left|\tilde{\Psi}_{n,l,\{\mu\}}\left( \vec{p} \right) \right|^2={\cal{M}}^2_{n,l}(p) \left[ {\cal{Y}}_{l\{\mu\}}(\Omega_{D-1})\right]^2
\label{denmom}
\eeq
for the quantum-mechanical probability density of the system in momentum space.

\section{The shape complexity of the D-dimensional hydrogenic system}\label{sec3}

Here we describe the methodology to compute the position and momentum shape complexity of our system in an arbitrary physical state characterized by the hyperquantum numbers $\left( \eta, \mu_1, ..., \mu_{D-1} \right)$. We show that the calculation of the position and momentum hydrogenic shape complexities ultimately reduce to the evaluation of some entropic functionals of Laguerre and Gegenbauer polynomials.

\subsection{Position space}

The shape complexity $C \left[\rho \right]$ of the position probability density $\rho\left(\vec{r}\right)$ is defined \cite{C02} as
\beq
C\left[ \rho \right]= \left\langle   \rho \right\rangle \exp \left( S\left[ \rho \right] \right),
\label{defCpos}
\eeq
where
\beq
\left\langle   \rho \right\rangle = \int \rho^2 \left(\vec{r} \right) d \vec{r}
\label{Drho}
\eeq
and
\beq
S\left[ \rho \right]= - \int \rho \left(\vec{r} \right)  \ln  \rho \left(\vec{r} \right)  d\vec{r}
\label{Srho}
\eeq
denote the first-order frequency moment (also called averaging density or disequilibrium, among other names) and the Shannon entropy of $\rho \left( \vec{r} \right)$, respectively. Then, this composite information-theoretic quantity measures the complexity of the system by means of a combined balance of the average height (as given by $ \left\langle \rho \right\rangle)$ and the total bulk extent (as given by $S \left[ \rho \right]$) of the corresponding quantum-mechanical probability density $\rho \left( \vec{r} \right)$.

Let us first calculate $\left\langle   \rho \right\rangle$. From (\ref{FunOndaR}) and (\ref{Drho}) one obtains that
\beq
\left\langle   \rho \right\rangle = \frac{2^{D-2}}{\eta^{D+2}} Z^D K_1 \left( D,\eta,L \right) K_2 \left( l, \left\lbrace \mu \right\rbrace \right),
\label{Drho1}
\eeq
where
\beq
K_1 \left( D,\eta,L \right) = \int_{0}^{\infty} x^{-D-5} \left\lbrace \omega_{2L+1} (x) \left[ {\tilde{L}}^{2L+1}_{\eta-L-1} (x) \right]^2  \right\rbrace^2 dx
\label{K1}
\eeq
and
\beq
K_2 \left( l, \left\lbrace \mu \right\rbrace \right) = \int_{\Omega} \left| {\cal{Y}}_{l\{\mu\}}\left(\Omega_{D-1}\right) \right|^4 d\Omega_{D-1}
\label{K2}
\eeq

The Shannon entropy of $\rho \left( \vec{r} \right)$ has been recently shown \cite{DL08} to have the following expression
\beq
S \left[ \rho \right]= S \left[ R_{nl} \right] + S \left[ {\cal{Y}}_{l\{\mu\}} \right],
\label{Srho1}
\eeq
with the radial part
\begin{align}\nonumber
S\left[ R_{n,l} \right]& = - \int_{0}^{\infty} r^{D-1} R_{n,l}^{2} (r) \log R_{n,l}^{2} dr \\\label{SR}
& = A(n,l,D) + \frac{1}{2 \eta} E_1\left[ \tilde{{\cal{L}}}^{2L+1}_{\eta-L-1} \right]-D \ln Z
\end{align}
and the angular part
\begin{align}\nonumber
S \left[{\cal{Y}}_{l,\left\lbrace \mu \right\rbrace} \right]& =- \int_{S_{D-1}} \left|{\cal{Y}}_{l,\left\lbrace \mu \right\rbrace} \left( \Omega_{D-1}\right)  \right|^2 \ln \left| {\cal{Y}}_{l,\left\lbrace \mu \right\rbrace} \left( \Omega^{'}_{D-1}\right)  \right|^2 d\Omega_{D-1} \\\label{SY}
& = B (l,\left\lbrace \mu \right\rbrace,D)+\sum^{D-2}_{j=1} E_2 \left[ \tilde{C}^{\alpha_j+\mu_{j+1}}_{\mu_j-\mu_{j+1}} \right],
\end{align}
where $A(n,l,D)$ and $B(l,\left\lbrace \mu \right\rbrace , D)$ have the following values 
\beq
A(n,l,D) = -2l \left[\frac{2\eta-2L-1}{2 \eta} +\psi (\eta+L+1) \right]+\frac{3 \eta^2-L(L+1)}{\eta}+\ln \left[ \frac{\eta^{D+1}}{2^{D-1}} \right]
\nonumber
\eeq
and
\beq
B (l,\left\lbrace \mu \right\rbrace,D)= \ln 2\pi -2 \sum^{D-2}_{j=1} \mu_{j+1} \left[\psi(2\alpha_j+\mu_j+\mu_{j+1})-\psi(\alpha_j+\mu_j)-\ln 2- \frac{1}{2 (\alpha_j+\mu_j)}\right]
\nonumber
\eeq
with $\psi(x) = \frac{\Gamma^{'}(x)}{\Gamma(x)}$ is the digamma function. The entropic functionals $E_i \left[ {\tilde{y}}_{n} \right]$, $i=1$ and $2$, of the polynomials $\left\lbrace  {\tilde{y}}_{n} \right\rbrace $, orthonormal with respect to the weight function $\omega(x)$, are defined \cite{D98, ADY} by
\beq
E_1 \left[{\tilde{y}}_n \right]=-\int_{0}^{\infty} x\hspace*{1mm} \omega(x)\hspace*{1mm} {\tilde{y}}^{2}_{n} (x) \ln {\tilde{y}}^{2}_{n}(x) dx,
\label{E1}
\eeq
and
\beq
E_2 \left[{\tilde{y}}_n\right]=-\int_{-1}^{+1} \omega(x) \hspace*{1mm}{\tilde{y}}^{2}_{n} (x) \ln {\tilde{y}}^{2}_{n}(x) dx,
\label{E2}
\eeq
respectively.

Finally, from Eqs. (\ref{defCpos}), (\ref{Drho1}) and (\ref{Srho1})-(\ref{SY}), we obtain the following value for the position shape complexity of our system:
\begin{align}
C \left[ \rho \right] &= \frac{2^{D-2}}{\eta^{D+2}} K_1 \left(D, \eta, L \right) K_2 \left( L, \left\lbrace \mu \right\rbrace \right)\nonumber \\
& \quad \times \exp \left[ A(n,l,D) + \frac{1}{2 \eta} E_1 \left[{\tilde{L}}_{\eta-L-1}^{2 L+1} \right] +S\left[ {\cal{Y}}_{l,\left\lbrace \mu \right\rbrace} \right]\right],\label{CposFinal}
\end{align}
where the entropy of the hyperspherical harmonics $S\left[ {\cal{Y}}_{l,\left\lbrace \mu \right\rbrace} \right]$, given by Eq. (\ref{SY}), is controlled by the entropy of Gegenbauer polynomials $E_2 \left[{\tilde{C}}_k^{\alpha}\right]$ defined by Eq. (\ref{E2}). It is important to remark that the position complexity $C \left[\rho \right]$ does not depend on  the strength of the Coulomb potential, that is, on the nuclear charge $Z$.

\subsection{Momentum space}

The shape complexity $C \left[ \gamma \right]$ of the momentum probability density $\gamma \left( \vec{p} \right)$ is given by
\beq
C\left[ \gamma \right]= \left\langle   \gamma \right\rangle \exp \left( S\left[ \gamma \right] \right),
\label{defCmom}
\eeq
where the momentum averaging density $\left\langle \gamma \right\rangle$ can be obtained from Eq. (\ref{denmom}) as follows:

\beq
\left\langle   \gamma \right\rangle = \int \gamma^2 \left(\vec{p} \right) d \vec{p}= \frac{2^{4L+8} \eta^{D} }{Z^D} K_3 \left( D,\eta,L \right) K_2 \left( l, \left\lbrace \mu \right\rbrace \right),
\label{Cmom}
\eeq
with $K_2$ is given by Eq. (\ref{K2}), and $K_3$ can be expressed as
\beq
K_3 \left( D,\eta,L \right) = \int_{0}^{\infty} \frac{y^{4l+D-1}}{(1+y^2)^{4L+8}} \left[ {\tilde{C}}^{L+1}_{\eta-L-1} (\frac{1-y^2}{1+y^2}) \right]^4  dy
\label{K3}
\eeq

On the other hand, the momentum Shannon entropy $S\left[ \gamma \right]$ can be calculated in a similar way as in the position case. We have obtained that
\begin{align}\nonumber
S \left[ \gamma \right] & = - \int \gamma \left( \vec{p} \right) \ln \gamma \left( \vec{p} \right) d\vec{p} = S \left[ {\cal{M}}_{nl} \right] + S \left[ {\cal{Y}}_{l,\{\mu\}} \right] \\ \label{Sgamma}
&= F \left(n,l,D \right)+ E_2 \left[{\tilde{C}}_{\eta-L-1}^{L+1} \right]+D \ln Z+S\left[ {\cal{Y}}_{l,\left\lbrace \mu \right\rbrace} \right],
\end{align}
where $F\left(n,l,D \right)$ has been found to have the value
\begin{align}\label{F1}
F\left(n,l,D\right)&=-\ln \frac{\eta^D}{2^{2L+4}}-(2L+4) \left[ \psi(\eta+L+1)-\psi (\eta) \right]\nonumber\\
&\quad +\frac{L+2}{\eta}-(D+1) \left[1-\frac{2 \eta (2L+1)}{4\eta^2-1} \right]
\end{align}

Then, from Eqs. (\ref{defCmom}), (\ref{Cmom}) and (\ref{Sgamma}) we finally have the following value for the momentum shape complexity
\begin{align}
C \left[ \gamma \right] &= 2^{4 L+8} \eta^D  K_3 \left(D, \eta, L \right) K_2 \left( L, \left\lbrace \mu \right\rbrace \right)\nonumber \\ 
& \quad \times \exp \left\lbrace  F(n,l,D) + E_2 \left[{\tilde{C}}_{\eta-L-1}^{L+1} \right] +S\left[ {\cal{Y}}_{l,\left\lbrace \mu \right\rbrace} \right] \right\rbrace \label{CmomFinal}
\end{align}

Notice that, here again, this momentum quantity does not not depend on the nuclear charge Z. Moreover the momentum complexity $C \left[\rho \right]$ is essentially controlled by the entropy of the Gegenbauer polynomials $E_2 \left[ {\tilde{C}}_{k}^{\alpha} \right]$, since the entropy of hyperspherical harmonics $S\left[ {\cal{Y}}_{l,\left\lbrace \mu \right\rbrace} \right]$ reduces to that of these polynomials according to Eq. (\ref{SY}).

\section{Shape complexities of ground and circular states}\label{sec4}

Here we apply the general expressions (\ref{CposFinal}) and (\ref{CmomFinal}) found for the position and momentum shapes complexities of an arbitrary physical state of the D-dimensional hydrogenic system, respectively, to the ground state $\left( n=1,\mu_i=0, \forall i=1...D-1 \right)$ and to the circular states. A circular state is a single-electron state with the highest hyperangular momenta allowed within a given electronic manifold, i.e. a state with hyperangular momentum quantum numbers $\mu_i=n-1$ for all $i=1,...,D-1$.

\subsection{Ground state}

In this case $\eta-L-1=0$, so that the Laguerre polynomial involved in the radial wavefunction is a constant. Then, the probability density of the ground state in position space given by Eqs. (\ref{radial}), (\ref{hyperesf}) and (\ref{probdif}) reduces as follows:
\beq
\rho_{g.s.}(\vec{r})= \left(\frac{2 Z}{D-1} \right)^D \frac{1 }{\pi^{\frac{D-1}{2}} \Gamma\left( \frac{D+1}{2} \right) } e^{-\frac{4Z}{D-1}r},
\label{denPosGS}
\eeq
which has been also found by various authors (see e.g. \cite{H93, A00}).

The expressions (\ref{Drho1})-(\ref{K2}), which provide the averaging density of arbitrary quantum-mechanical state, reduce to the value
\beq
\left\langle \rho_{g.s.} \right\rangle = \frac{Z^D}{(D-1)^D} \frac{1}{\pi^{\frac{D-1}{2}} \Gamma \left( \frac{D+1}{2} \right)}
\label{DposGS}
\eeq
for the ground-state averaging density. Moreover, the angular part of the entropy is
\beq
S \left[{\cal{Y}}_{0,\left\lbrace 0 \right\rbrace  } \right]= \ln \frac{2 \pi^{D/2}}{\Gamma \left( \frac{D}{2} \right)},
\label{SYgs}
\eeq
so that it is equal to  $\ln 2 \pi$ and $\ln 4 \pi$ for $D=2$ and $3$, respectively. Then, the formulas (\ref{Srho1})-(\ref{E2}) of the Shannon entropy of arbitrary physical state of our system simplify as
\beq
S \left[ \rho_{g.s.} \right]=\ln \left( \frac{(D-1)^D}{2^D} \pi^{\frac{D-1}{2}} \Gamma \left( \frac{D+1}{2} \right)  \right)+D-D \ln Z 
\label{SposGS}
\eeq
for the ground-state Shannon entropy. Finally, from Eq. (\ref{CposFinal}) or from its own definition together with (\ref{DposGS})-(\ref{SposGS}) we obtain that the position shape complexity of D-dimensional hydrogenic ground state has the value
\beq
C \left[\rho_{g.s.} \right]= \left( \frac{e}{2}\right)^D
\label{CposGS}
\eeq

In momentum space we can operate in a similar manner. First we have seen that the ground-state probability density is
\beq
\gamma_{g.s.}(\vec{p})= \frac{(D-1)^D \Gamma \left(\frac{D+1}{2} \right) }{Z^D \pi^{\frac{D+1}{2}}} \frac{1}{\left( 1+\frac{(D-1)^2}{4} {\tilde{p}}^2 \right)^{D+1}},
\label{denMomGS}
\eeq
which has been also given by Aquilanti et al \cite{AQ}, among others. Then, we have found the values
\beq
\left\langle \gamma_{g.s.} \right\rangle = \left( \frac{2D-2}{Z} \right)^D \frac{1}{\pi^{\frac{D+2}{2}}} \frac{\Gamma^2 \left( \frac{D+1}{2} \right) \Gamma \left(2+\frac{3D}{2} \right)}{\Gamma \left(2D+2 \right)}
\label{DmomGS}
\eeq
for the momentum averaging density, and
\beq
S \left[ \gamma_{g.s.} \right]=\ln \frac{\pi^{\frac{D+1}{2}}}{(D-1)^{D} \Gamma\left(\frac{D+1}{2} \right) } + (D+1) \left[ \psi (D+1)-\psi \left( \frac{D}{2}+1 \right)  \right]+D \ln Z
\label{SmomGS}
\eeq
for the momentum Shannon entropy, directly from Eq. (\ref{DposGS}) or from Eqs. (\ref{Cmom})-(\ref{K3}) and (\ref{Sgamma})-(\ref{F1}), respectively. Finally. from Eq. (\ref{CmomFinal}) or by means of Eqs. (\ref{DmomGS})-(\ref{SmomGS}) we have the following value
\beq
C \left[\gamma_{g.s.} \right]= \frac{2^D \Gamma \left( \frac{D+1}{2}\right) \Gamma \left(2+\frac{3 D}{2} \right) } {\pi^{1/2} \Gamma \left( 2D+2\right)} \exp \left\lbrace  (D+1) \left[ \psi \left( D+1 \right)-\psi \left( \frac{D+2}{2} \right)  \right]  \right\rbrace 
\label{CmomGS}
\eeq
for the ground-state D-dimensional hydrogenic shape complexity in momentum space. In particular, this quantity has the values
\beq
C_2 (\gamma_{g.s.})= \frac{2 e^{3/2}}{5}=1.7926
\nonumber
\eeq
\beq
C_3 (\gamma_{g.s.})=\frac{66}{e^{10/3}}=2.3545
\nonumber
\eeq
\beq
C_4 (\gamma_{g.s.})=\frac{e^{35/12}}{6}=3.0799
\nonumber
\eeq
for the hydrogenic system with dimensionalities $D=2,3$ and $4$, respectively. Let us here mention that the three-dimensional value agrees with that calculated in \cite{C02}.

\subsection{Circular states}

Following a parallel process with circular states, we have obtained
\beq
\rho_{c.s.}(\vec{r}) =\frac{2^{D+2-2n} Z^D  }{\pi^{\frac{D-1}{2}}(2n+D-3)^{D} \Gamma(n) \Gamma \left(n+\frac{D-1}{2} \right)  } e^{-\frac{r}{\lambda} } \left( \frac{r}{\lambda} \right)^{2n-2} \prod^{D-2}_{j=1} \left( \sin \theta_j \right)^{2n-2}
\nonumber
\eeq
for the position probability density, and
\beq
\gamma_{c.s.}(\vec{p})= \frac{2^{2n-2} (2n+D-3)^D \Gamma \left( n+\frac{D-1}{2}\right)}{Z^D \pi^{\frac{D+1}{2}} \Gamma(n)} \frac{(\eta p/Z)^{2n-2}}{(1+\frac{\eta^2 p^2}{Z^2})^{2n+D-1}} \prod^{D-2}_{j=1} \left( \sin \theta_j \right)^{2n-2}
\nonumber
\eeq
for the momentum probability density of a D-dimensional hydrogenic circular state with the principal quantum number $n$. Moreover, we have found the values
\beq
\left\langle \rho_{c.s.} \right\rangle = \frac{Z^D \Gamma \left(n-\frac{1}{2} \right) \Gamma \left(2 n + \frac{D-3}{2} \right)}{2^{2n-2} \pi^{\frac{D}{2}} (2n+D-3)^D  \Gamma \left(n \right) \Gamma^2 \left(n+\frac{D-1}{2} \right)}
\label{DposCS}
\eeq
and
\beq
\left\langle \gamma_{c.s.} \right\rangle = \frac{2^{4n+D-4} (2n+D-3)^D \Gamma^2 \left( n+\frac{D-1}{2} \right) \Gamma \left(2n-1 \right) \Gamma \left( 2n+\frac{3D}{2} \right)}{Z^D \pi^{\frac{D+2}{2}} \Gamma^2 \left(n \right) \Gamma \left( 4n+2D-2 \right)}
\label{DmomCS}
\eeq
for the position and momentum averaging densities of our system. On the other hand, we have also been able to express the position and momentum entropies as
\begin{align}\label{SposCS}
S \left[\rho_{c.s.} \right]&= 2n+D-2 -(n-1) \left[ \psi (n) + \psi \left(n+\frac{D-1}{2} \right)  \right] -D \ln 2\\\nonumber
&\qquad+\ln \left[(2n+D-3)^D \pi^{\frac{D-1}{2}} \Gamma(n) \Gamma  \left(n+\frac{D-1}{2} \right)\right]- D \ln Z
\end{align}
and
\beq\label{SmomCS}
S \left[\gamma_{c.s.} \right]= A(n,D)+\ln \left[ \frac{2^{D+1} Z^D \pi^{\frac{D+1}{2}} \Gamma(n) }{(2n+D-3)^D \Gamma  \left(n+\frac{D-1}{2} \right) }\right],
\eeq
where the constant $A(n,D)$ is given by
\begin{align}\nonumber
A(n,D)=&\frac{2n+D-1}{2n+D-3}-\frac{D+1}{2n+D-2}-(n-1) \psi (n) \\\label{A2}
&\quad -\left(\frac{D+1}{2}\right) \psi \left(n+\frac{D-2}{2} \right) +\left(n+\frac{D-1}{2}\right) \psi \left(n+\frac{D-3}{2} \right)
\end{align}

Finally, from Eqs. (\ref{DposCS})-(\ref{SmomCS}) or from Eqs. (\ref{CposFinal}) and (\ref{CmomFinal}) we have the values
\begin{align}\nonumber
C \left[ \rho_{c.s.} \right] & = \frac{\Gamma  \left(n-\frac{1}{2} \right) \Gamma \left(2n+\frac{D-3}{2} \right)}{2^{2n+D-2} \pi^{1/2} \Gamma \left(n + \frac{D-1}{2} \right)}\\ \label{CposCS}
& \quad \times \exp \left\lbrace 2n+D-2-(n-1) \left[ \psi(n) + \psi \left(n+ \frac{D-1}{2} \right) \right] \right\rbrace
\end{align}
and
\beq
C \left[ \gamma_{c.s.} \right]= \frac{2^{4n+2D-3}\Gamma \left(n+ \frac{D-1}{2} \right) \Gamma(2n-1) \Gamma \left(2 n + \frac{3D}{2} \right)}{\pi^{1/2} \Gamma(n) \Gamma(4n+2D-2)} \exp \left[ A(n,D) \right]
\label{CmomCS}
\eeq
for the position and momentum shape complexity of a D-dimensional hydrogenic system in an arbitrary circular state. It is worthy to remark for checking purposes that Eqs. (\ref{CposCS}) and (\ref{CmomCS}) reduce to Eqs. (\ref{CposGS}) and (\ref{CmomGS}) in case that $n=1$, respectively, as expected; in this sense we have to use the two following properties of the digamma function: $\psi \left(  2 z \right)= \frac{1}{2} \left[ \psi \left(z \right)+  \psi \left(z +\frac{1}{2}\right) \right]+ \ln 2 $ and $\psi \left(  z+1 \right)=  \psi \left(z \right)+ \frac{1}{z}$.

\section{Conclusion}

The shape complexity of the hydrogenic system in D-dimensional position and momentum spaces is investigated. This quantity has two information-theoretic ingredients: the disequilibrium and the Shannon entropic power. We have seen that the explicit computation of this complexity is a formidable open task, mainly because the analytical evaluation of the entropic functionals of the Laguerre and Gegenbauer polynomials, $E_1 \left[{\tilde{L}}_{k}^{\alpha} \right]$ and $E_2 \left[{\tilde{C}}_{k}^{\alpha} \right]$, involved in the calculation of the Shannon entropy, has not yet been accomplished.

The general methodology presented here is used to find explicit expressions for the position and momentum complexities of the ground and circular states in terms of the dimensionality and the principal quantum number.


The authors gratefully acknowledge the Spanish MICINN grant FIS2008-02380 and the grants FQM-481, 1735 and 2445 of the Junta de Andaluc\'ia. They belong to the Andalusian research group FQM-207. S.L.R. and D.M acknwoledge the corresponding FPU and FPI scholarships of the Spanish Ministerio de Ciencia e Innovaci\'{o}n, respectively.


\begin{thebibliography}{99}

\bibitem{H93}{ D. R. Herschbach, J. Avery and O. Goscinski, {\it Dimensional Scaling in Chemical Physics} (Kluwer, Dordrecht, 1993). See Chapter 5.}

\bibitem{A00} {J. Avery, {\it Hyperspherical Harmonics and Generalized Sturmians} (Kluwer, Dordrecht,2000)}

\bibitem{H05} {P. Harrison, {\it Quantum Wells, Wires and Dots: Theoretical and Computational Physics of Semiconductor Nanostructures} (Wiley-Interscience, New York, 2005). Second edition}

\bibitem{L07} {S.S. Li and J.B. Xia, Phys. Lett. A \textbf{366} (1-2) (2007) 120-123. See references herein.}

\bibitem{N00}{ M.M. Nieto, Amer. J. Phys. \textbf{47} (1979) 1067 }

\bibitem{D03}{ M.I. Dykman, P.M. Platzman and P. Seddigard, Phys. Rev. B \textbf{67} (2003) 155402}

\bibitem{I06}{C. Itzykson and J.B. Zuber, {\it Quantum Field Theory} (Dover, 2006)}

\bibitem{B99}{F. Burgbacher, C. L\"{a}mmerzahl and A. Macias,  J. Math. Phys. \textbf{40} (1999) 625-634 958 242862}

\bibitem{Y94}{ R.J. Y\'{a}\~{n}ez, W. Van Assche and J.S. Dehesa, Phys. Rev. A \textbf{50} (1994), 3065-3079}

\bibitem{D98}{ J.S. Dehesa, R.J. Y\'{a}\~{n}ez, A.I. Aptekarev and V. Buyarov,  J. Math. Phys. \textbf{39} (6) (1998) 3050-3060}

\bibitem{D01}{ J.S. Dehesa, A. Mart\'{i}nez-Finkelshtein and J. S\'{a}nchez-Ruiz,  J. Comp. Appl. Math. \textbf{133} (2001) 23-46}

\bibitem{V00}{ W. van Assche, R.J. Y\'{a}\~{n}ez, R. Gonz\'{a}lez-F\'{e}rez and J.S. Dehesa,  J. Math. Phys. \textbf{41} (2000) 6600-6613}


\bibitem{D08}{ J.S. Dehesa, S. Lopez-Rosa, A. Mart\'inez-Finkelshtein and R.J. Y\'a\~nez. Preprint 2009.}

\bibitem{S08}{ J. Sa\~nudo and R. Lopez-Ruiz, Physics Letters A \textbf{372} (2008) 5283-5286.}

\bibitem{L08}{ S. Lopez-Rosa, J.C. Angulo and J. Antolin, Physica A (2009). In press.}

\bibitem{C02}{ R.G. Catalan, J. Garay and R. Lopez-Ruiz, Phys. Rev. E \textbf{66} (2002) 011102.}

\bibitem{A93}{ J. Avery, J. Phys. Chem. \textbf{97} (1993) 2406-2412.}

\bibitem{DL07}{ J.S. Dehesa, S. L\'opez-Rosa, B. Olmos and R.J Y\'a\~nez,  J. Math. Phys. \textbf{47} (2006) 052104}

\bibitem{AQ}{ V. Aquilanti, S. Cavalli and C. Coletti, Chem. Phys. \textbf{214} (1997) 1-13}

\bibitem{DL08}{ J.S. Dehesa, S. L\'opez-Rosa, A. Mart\'inez-Finkelshteins and R.J Y\'a\~nez, in Proc. ECMI 2008 (London, 30 June-4 July 2008), to appear in Lecture Notes in Math.}

\bibitem{ADY}{ A.I. Aptekarev, J.S. Dehesa and R.J. Y\'a\~nez,, J. Math. Phys. \textbf{35} (1994) 4423}

\end{thebibliography}
\end{document}